\documentclass[aps,prl,reprint,groupedaddress,showpacs]{revtex4-1}

\usepackage{graphicx}
\usepackage{textcomp}
\usepackage{hyperref}
\usepackage{amssymb}
\usepackage{amsmath}
\usepackage{ifthen} 
\usepackage{xifthen} 

\newcommand{\eq}[1]{equation~(\ref{#1})}
\newcommand{\fig}[2][]{%
\ifthenelse{\isempty{#1}}
{Fig.~\ref{#2}}
{Fig.~\ref{#2}(#1)}
}

\begin{document}


\title{Frequency and Q-factor control of nanomechanical resonators}


\author{Johannes Rieger}
\author{Thomas Faust}
\author{Maximilian J. Seitner}
\author{J\"org P. Kotthaus}
\author{Eva M. Weig}
\email[]{weig@lmu.de}
\affiliation{Center for NanoScience (CeNS) and Fakult\"at f\"ur Physik, Ludwig-Maximilians-Universit\"at, Geschwister-Scholl-Platz 1,
M\"unchen 80539, Germany}



\begin{abstract}
We present an integrated scheme for dielectric drive and read-out of high-Q nanomechanical resonators which enables tuning of both the resonance frequency and quality factor with an applied DC voltage.
A simple model for altering these quantities is derived, incorporating the resonator's complex electric polarizability and position in an inhomogeneous electric field, which agrees very well with the experimental findings as well as FEM simulations.
By comparing two sample geometries we are able to show that careful electrode design can determine the direction of frequency tuning of flexural in- and out-of-plane modes of a string resonator. 
Furthermore we demonstrate that the mechanical quality factor can be voltage reduced more than fivefold.
\end{abstract}


\maketitle

Control of small-scale mechanical systems is essential for their application. Resonant micro- and nanoelectromechanical systems (M/NEMS) have both proven themselves technologically viable (frequency filtering in cell phones \cite{Bouchaud2005}, gyroscopes \cite{Dean2009}, atomic force microscope (AFM) cantilevers\cite{Binnig1986}) as well as shown great promise for next-generation sensor applications (mass sensors \cite{Naik2009,Li2010,ChasteJ.2012}, resonant bio sensors \cite{Burg2007} and ultra sensitive force sensors \cite{Mamin2001,Regal2008}).
Three areas of development are central to realizing the potential of high performance resonant micro- and nanomechanics: advancement of high Q geometries and materials; improved readout schemes for mechanical motion, including compactness and integrability; and increased control of the resonant behavior of the mechanics.
In the field of nanomechanics, the last years have seen the advent of high Q silicon nitride strings under high tensile stress \cite{Verbridge2006}, as well as indications that the stress rather than the nitride is responsible for the high mechanical quality factor \cite{Unterreithmeier2010}. 
Subsequently, efficient integrated drive and read-out schemes have been developed to detect the sub-nanoscale motion of small-scale resonant mechanics\cite{Ekinci2005a}. 
Very good tunability of the resonance frequency can be achieved by capacitive coupling of the nanomechanical element to a side electrode \cite{Kozinsky2006}.
However, the required metalization of the resonant structure reduces the room temperature quality factor significantly\cite{Yu2012} via Ohmic losses.
In our lab, an efficient, room-temperature microwave mixing scheme has been developed for readout \cite{Faust2012}, as well as a dielectric drive mechanism to actuate mechanics regardless of their material make-up\cite{Unterreithmeier2009}, importantly obviating the necessity to metallize otherwise low-loss dielectrics.
In this paper we present a continuation of this development that enables tuning of both the frequency and quality factor of nanomechanical resonators in the context of this highly applicable and integrable scheme.
Using the combined dielectric actuation and microwave readout schemes, we theoretically develop the means to controllably raise and lower the resonant frequency of various flexural modes of our mechanics, as well as to broaden the mechanical resonance linewidth.
The latter so-called Q control \cite{Rodriguez2003,Sulchek2000}is widely used in AFM measurements and grants faster image acquisition rates by decreasing the mechanical response time.
In the presented scheme this Q control is achieved by simply applying a DC voltage instead of the normally employed active feedback mechanism and is thus well suited for integration.
The theoretical relationship between the design of the electrodes and the resulting control of a given mode is validated both by experiment and simulation.

\begin{figure}
\includegraphics{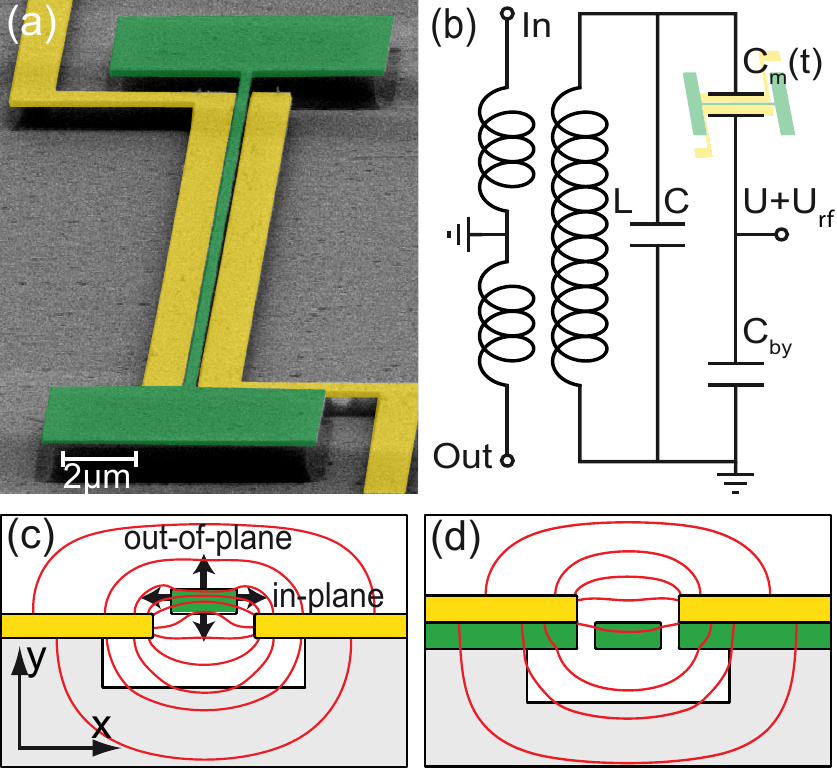}
\caption{\label{1}(color online). {\bf a)} False color SEM micrograph of a 55\,\textmu m long silicon nitride resonator in the configuration depicted in c). 
The gold electrodes forming the capacitor $\rm{C_m(t)}$ are highlighted in yellow, the silicon nitride resonator and its clamping points in green.
{\bf b)} Equivalent circuit diagram of the transduction scheme with an inductively-coupled microwave cavity - represented by the capacitance C and inductance L - for dielectric readout. $\rm{C_m(t)}$ is the capacitance between the gold electrodes which is modulated by resonator displacement.
The microwave bypass capacitor $\mathrm{C_{by}}$ allows the additional application of a DC and RF voltage, U and $\rm{U_{rf}}$ respectively, and thus dielectric tuning and driving. 
{\bf c,d)} Schematic cross section with simulated field lines for the elevated and lowered geometry. The cross section of the silicon nitride beam is 260\,nm x 100\,nm. The arrows in c) describe the directions of the in-plane and out-of-plane oscillation.}
\end{figure}

Our system is depicted in \fig{1}.
A nanomechanical silicon nitride string is situated between a pair of near-lying electrodes (\fig[a]{1}).
A DC voltage across the electrodes produces an electric field that induces a dipolar moment in the silicon nitride string, which thus experiences a force in the gradient field.
This force can be modulated by an additional AC voltage, actuating the mechanical resonance\cite{Unterreithmeier2009}.
Additionally, the mechanical resonator is dielectrically coupled to an external microwave cavity via the same electrodes.
An equivalent circuit diagram is shown in \fig[b]{1}.
Deflection of the beam translates into a change of the capacitance $\rm{C_m(t)}$ between the two electrodes and thereby modulates the cavity transmission signal.
The mechanical oscillation can then be detected by demodulating this signal.
A more detailed description of this room temperature heterodyne detection including the demonstration of self-oscillation caused by cavity backaction can be found elsewhere \cite{Faust2012}.

To enable direct actuation of the mechanical resonator, we introduce a microwave bypass between ground and one of the electrodes using the single layer capacitor (SLC)\footnote{500U04A182KT4S from Johanson Technology} $\rm{C_{by}}$.
Thus a DC and RF voltage can be applied to this electrode, whereas the other electrode is grounded via the microstrip cavity (compare \fig[b]{1}).
Consequently, the presented integrated configuration combines the above mentioned cavity-enhanced detection\cite{Faust2012} with dielectric driving and tuning \cite{Unterreithmeier2009}. 
Note that the drive and detection can be realized with a single set of electrodes, since crosstalk is avoided by operating the cavity at 3.5\,GHz, well-separated in frequency from the mechanical fundamental flexural mode frequency of here 6.5\,MHz.  


\begin{figure*}[htb]
\includegraphics{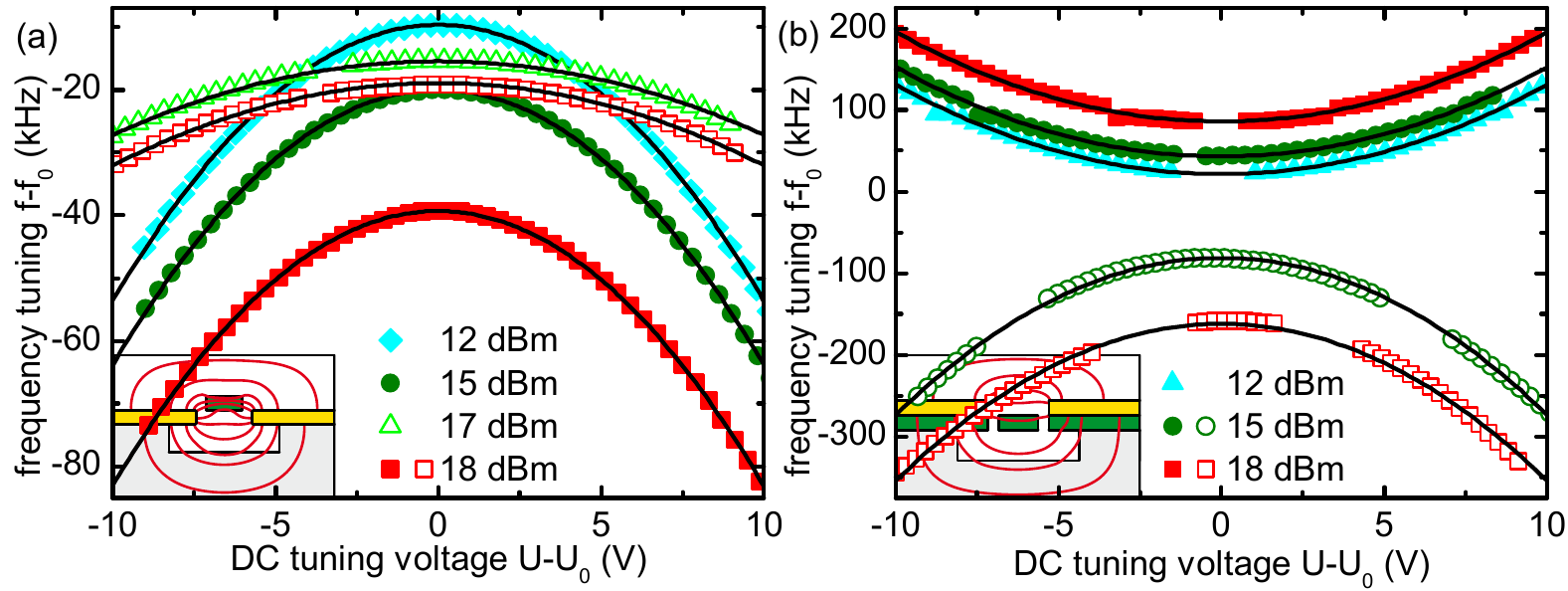}
\caption{\label{2}(color online). Quadratic tuning of the mechanical resonance frequency with DC voltage $U$ for the two different geometries. 
The graphs show the deviation of the resonance frequency $f$ from the natural resonance frequency $f_0$ of the resonator's respective mode (in- or out-of-plane, depicted as open and filled symbols respectively) for different microwave cavity pump powers indicated in dBm. The solid lines are a fit of the model.
{\bf (a)} The force gradient has the same parity for the in-plane- (open symbols, small curvature) as well as the out-of-plane mode (filled symbols, large curvature). {\bf (b)} Tuning direction differs for the two modes. The out-of-plane mode tunes upwards, the in-plane-mode downwards in frequency.}
\end{figure*}

For this study, two sample geometries for obtaining optimized gradient field coupling are fabricated from high-stress silicon nitride using standard electron beam lithography, reactive ion etching and subsequent hydrofluoric wet etching to obtain suspended strings.
To minimize Ohmic losses of the microwave circuit originating in the bulk chip supporting the mechanical resonator, we use a fused silica substrate instead of silicon.
The geometries are schematically shown in \fig[c,d]{1}. 
Referring to the beam's position with respect to the electrodes, the two structures will from now on be referenced as ``elevated'' (\fig[c]{1}) and ``lowered'' (\fig[d]{1}).
The centerpiece of each structure is the 55\,\textmu m long silicon nitride string resonator with a rectangular cross section of width 260\,nm and height 100\,nm. 
The freely suspended resonator is bordered by two vertically offset gold electrodes, one of which is connected to the microstrip cavity with a resonance frequency of 3.5\,GHz and a quality factor of 70, while the other electrode leads to the SLC. 
The essential difference between the geometries is the vertical positioning of the beam with respect to the gold electrodes. 
This affects the dielectric environment and thereby the electric field lines as depicted in \fig[c,d]{1}.
In the elevated design shown in \fig[c]{1}, the upper edge of the electrodes coincides with the lower edge of the beam, whereas in the lowered design in \fig[d]{1}, the upper edge of the beam coincides with the lower edge of the electrodes.
The simulated electric field lines for both geometries are obtained from finite element simulations using \textit{COMSOL Multiphysics} and allow us to extract the electric field along the x- and y-direction.
These inhomogeneous electric fields cause force gradients for the in- and out-of-plane modes of the resonator.
They thus alter the restoring force of the respective mode and thereby its resonance frequency \cite{Unterreithmeier2009}.
At the same time, the mechanical quality factor can be altered with the DC voltage, as the strong electric field and high field gradient lead to velocity-dependent dielectric losses in the beam material.
This frequency and linewidth tuning can be described by a simple model which agrees very well with our experimental findings and finite element simulations.
The resonance frequency can be tuned over 5\,\% and the resonance linewidth can be increased by a factor of 6.

We find the force gradient to be proportional to the square of the voltage and thus expect a quadratic dependence of the resonator resonance frequency on the applied DC voltage.
This can be derived from the energy of the induced dipolar moment $\vec{p}=\alpha\vec{E}$ of the dielectric resonator in an external electric field $\vec{E}$, where $\alpha$ is the polarizability of the silicon nitride beam. 
Assuming that $\vec{p}$ and $\vec{E}$ are parallel, i.\,e. using a scalar, complex polarizability $\alpha=\alpha'+i\alpha''$ and introducing a dependence of the electric field on the variable coordinate $\xi$, the energy $W$ reads 

\begin{equation}
W=\vec{p}\cdot\vec{E}=pE=\alpha E^2=(E(\xi))^2(\alpha'+i\alpha'')
\end{equation}

Here $\xi$ can be the x- or y-coordinate (compare \fig[c]{1}), so the following considerations apply to both the in- and out-of-plane mode.
Assuming $E(\xi)=E_0+E_1\xi$ for small displacements, the total energy can be separated into a real (stored) and an imaginary (dissipative) part.
\begin{eqnarray}
W_{\rm stored}=\alpha'\left(E_0^2+2E_0E_1\xi+E_1^2\xi^2\right)\\
\label{wloss}
W_{\rm loss}=\alpha''\left(E_0^2+2E_0E_1\xi+E_1^2\xi^2\right)
\end{eqnarray}

The stored energy $W_{\rm stored}$ is related to the frequency shift of the mechanical resonator, as the second derivative of this energy provides an additional force gradient i.\,e. an electrically induced spring constant $k_e$:
\begin{equation}
\label{ke}
k_e=-\frac{\partial F_e}{\partial \xi}=\frac{\partial^2W_{\rm stored}}{\partial \xi^2}=\alpha'E_1^2
\end{equation}
Hence the shift in resonance frequency caused by the electrical spring constant $k_e$ can be expressed in experimentally accessible units as
\begin{equation}
\label{fke}
f=\sqrt{\frac{k_0+k_e}{m}}\approx f_0+\frac{k_e}{2mf_0}=f_0+\frac{\beta^2U^2\alpha'}{2mf_0}
\end{equation}
which assumes both, the electrical frequency shift $k_e/2mf_0$ to be much smaller than the mechanical frequency $f_0$ and a geometry-dependent proportionality between the applied voltage and the resulting electric field gradient $E_1=\beta U$.
Moreover, as predicted by our finite-element simulations the sign of the gradient (corresponding to the sign of $\beta$) depends on the chosen geometry such that the out-of-plane mode changes its tuning direction between the elevated and the lowered design, which does not occur for the in-plane modes.

The quadratic tuning behavior with DC voltage is found to agree very well with the experimental data, as displayed in \fig{2}. 
All measurements are conducted at room temperature and a pressure of $10^{-4}$\,mbar.
For each mode and geometry the mechanical spectrum of both modes is taken for different DC voltages and microwave powers.
The driving voltage $U_{\rm rf}$ is kept constant in every measurement.
The values for $U_{\rm rf}$ lie within 80\,\,\textmu V and 1\,mV depending on the particular mode and geometry.

A Lorentzian fit to each mechanical spectrum yields the resonance frequency and the quality factor of the resonance for each parameter set.
The resonance frequencies lie around 6.5\,MHz and the highest quality factor is $340\,000$ for the out-of-plane mode in the elevated design.
Note that the tuning with microwave power is a result of the effective microwave voltage \cite{Faust2012}, and so is analogous to the tuning with a DC voltage. 
Subsequently, we fit $f=f_0 + c (U-U_0)^2 + c_{\rm{mw}} U_{\rm{mw}}^2$ to the tuning curves shown in \fig{2}, using the natural resonance frequency $f_0$ and two tuning parameters for the DC voltage $U$ and the effective microwave voltage ${U_{\rm mw}}$, as the DC and high frequency polarizability might differ.
We also introduce the DC offset $U_0$ to account for a shift (typically less than 1\,V) of the vertex of the tuning parabola away from zero DC voltage, which is most likely caused by trapped charges in the dielectric resonator material. 
As the influence of the microwave field on static dipoles averages out, there is no such shift resulting from the microwave voltage ${U_{\rm mw}}$.
Consequently, we can extract the tuning parameters for each geometry and oscillation direction.
With increasing voltage $U$ and for the elevated geometry depicted in \fig[a]{2}, both the in- and out-of-plane mode tune to lower frequencies, whereas for the lowered design (\fig[b]{2}), the out-of-plane mode tunes to higher frequencies, as predicted by our simulations.
The solid black lines in \fig{2} show the fit of our model with a single set of parameters for each mode in excellent agreement with the data. 
In the case of opposite frequency tuning, the initial frequency difference of the in- and out-of-plane modes can be evened-out, which leads to an avoided crossing caused by a coupling between the modes \cite{Faust2012a}. 
As the data points in this coupling region deviate from normal tuning behavior, they have been omitted in \fig[b]{2}.

Altering the DC or effective microwave voltage does not only shift the resonance frequency, but also influences the damping $\Gamma=2\pi f/Q=2\pi\Delta f$ of the mechanical resonance and thereby the measured linewidth $\Delta f$. 
Like the resonance frequency, the damping $\Gamma$ varies quadratically with increasing voltage. This can be understood by analyzing the dissipated energy ${W_{\rm loss}}$ given by \eq{wloss}: A time average of this quantity over one period of mechanical vibration $\xi(t)=\xi_0\cos(\omega t)$ gives

\begin{eqnarray}\nonumber
\label{wlossav}
\overline{W_{\rm loss}}&=& \frac{\alpha''}{T}\int\limits_0^T\left[2E_0E_1\xi_0\cos(\omega t) + E_1^2\xi_0^2\cos^2(\omega t)\right]dt \\
&& \left. = \frac{1}{2}\alpha''E_1^2\xi_o^2 \right.
\end{eqnarray}

Here we omit the $E_0^2$ term (as $\alpha''(\omega=0)=0$, otherwise static electric fields would lead to dissipation).
As the mechanical stored energy $W_{\rm mech}=\frac{1}{2}m\omega_0^2\xi_0^2$ is much larger than the electrical energy $W_{\rm stored}$, one can approximate the additional electrical damping to be

\begin{equation}
\label{gamma}
\Gamma_e(U)=\frac{\overline{W_{\rm loss}}\omega_0}{2\pi W_{\rm mech}}=\frac{\beta^2U^2\alpha''}{2\pi m\omega_0}
\end{equation}

\begin{figure}[htb]
\includegraphics{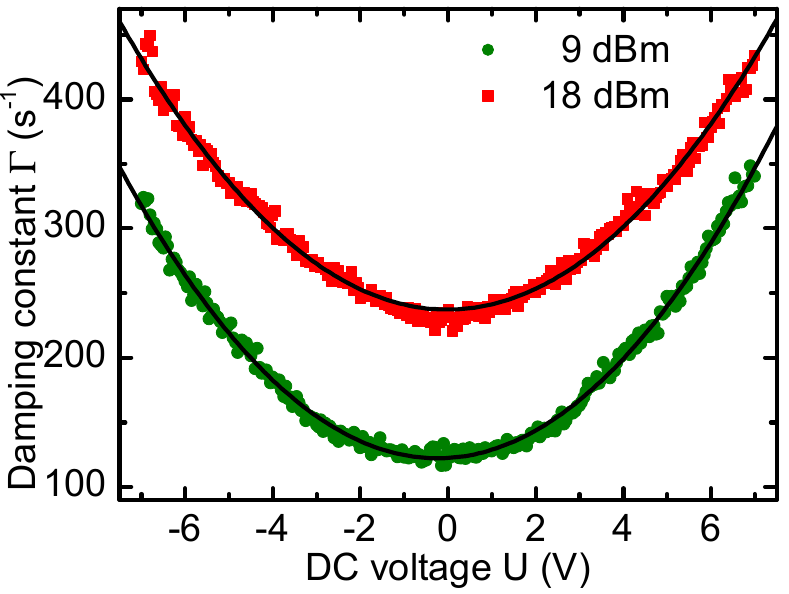}
\caption{\label{3}(color online). Damping constant versus DC voltage for two different microwave powers, exhibiting a quadratic behavior. The solid lines are a fit of the model mentioned in the text.}
\end{figure}

%

The measured damping versus DC voltage is shown in \fig{3}. It displays the quadratic behavior of the damping constant $\Gamma=\Gamma_0 + \Gamma_e(U)=\Gamma_0+c_\Gamma U^2$ of the out-of-plane mode in the elevated design for two different microwave powers. 
Here $\Gamma_0$ is the intrinsic damping of the resonator and $\Gamma_e(U)$ is given by \eq{gamma}.
Again, the vertical offset between the two curves is explained by the effective microwave voltage acting analogously to a DC voltage. 
The solid lines in \fig{3} are a fit of the model to the data, from which the curvature $c_\Gamma$ can be extracted.

Using this curvature and \eq{gamma}, the imaginary part of the polarizability can be expressed by
\begin{equation}
\alpha''=\frac{2\pi c_\Gamma m\omega_0}{\beta^2}
\end{equation}
Similarly, employing the curvature $c$ of the parabolic frequency shift and using \eq{fke} an expression for the real part $\alpha'$ can be given:
\begin{equation}
\label{alpha1}
\alpha'=\frac{2c m f_0}{\beta^2}
\end{equation}

The ratio 

\begin{equation}
\frac{\alpha''}{\alpha'}=\tan(\phi)=\frac{c_\Gamma}{2 c}
\end{equation}

is then independent of all resonator parameters and can be determined from the two curvatures.

The measured values for damping and tuning curvatures are $c_\Gamma=5.2\,\rm{\frac{1}{V^2s}}$ and $c=438\,\rm{\frac{Hz}{V^2}}$, leading to $\tan(\phi)=0.037$.
By using the Clausius-Mossotti-Relation
to first calculate the (lossless) $\alpha$ using $\epsilon=7.5$, one can determine the dielectric loss tangent

\begin{equation}
\tan(\delta)=\frac{\epsilon''}{\epsilon'}=0.016
\end{equation}

This value is well within the range of loss tangents reported for silicon nitride thin films\,\cite{Gould2003}.
Note that the time-varying capacitance $\rm{C_m(t)}$ induces a dissipative current in the electrodes, which also leads to a quadratically increasing damping\cite{Kozinsky2006}. 
However, using values obtained from FEM simulations for the electrode capacitance and its variation with beam deflection\cite{Unterreithmeier2009}, we estimate that this damping is three orders of magnitude smaller than that caused by dielectric losses. 
The relevant effect for the additional damping with increasing DC voltage is thus the dissipative reorientation of the dipoles in the resonator caused by its motion in a static, inhomogenous electric field, as described by \eq{wlossav}.

The DC voltage dependence of the mechanical damping $\Gamma$ was also measured at zero microwave power using an optical detection technique\,\cite{Kouh2005}.
The resulting $\Gamma_0$ was within a few percent of the value extracted from the 9\,dBm curve in \fig{3}, demonstrating that a measurement at low microwave powers induces only negligible additional damping to the mechanical resonator.
 
In conclusion, we show dielectric frequency tuning of over 5\,\% of the natural resonance frequency for nanomechanical resonators in an all-integrated setup that requires no metallization of the resonant mechanical structure itself.
This scheme thus maintains an excellent quality factor of up to $340\,000$ at $\rm{6.5\,MHz}$ and $300\,K$.
Furthermore, by careful design of the geometry, one can choose the tuning behavior of the out-of-plane mode to be either upward or downward in frequency and thus tune the two orthogonal resonator modes both in the same and in opposite directions.
We demonstrate that dielectric losses become highly relevant when using nanoscale electrode geometries generating large field gradients providing high tunability.
This could be used as a Q control \cite{Rodriguez2003,Sulchek2000,Venstra2011} that does not require any active electronics such as a phase-locked loop but rather a single DC voltage. 
Such a Q control can be employed to increase the bandwidth of NEMS sensors significantly, leading to much more adaptable devices.
Without the need for active electronics this could prove to be very well suited for integrated designs.
Together with microwave cavity backaction \cite{Faust2012} the mechanical resonance linewidth can be controlled from a few Hz up to more than 100\,Hz, thereby tuning the mechanical bandwidth by about two orders of magnitude. 
Finally, we imagine that the scheme presented can also be employed to build self-sensing AFM cantilevers\cite{Li2007} with tunable bandwidth and resonance frequency that are not subject to the bandwidth limitations of the normally employed piezo drive and could thus be used in multifrequency force microscopy schemes \cite{Garcia2012}.

\begin{acknowledgments}
Financial support by the Deutsche Forschungsgemeinschaft via Project No. Ko 416/18, the German Excellence Initiative via the Nanosystems Initiative Munich (NIM) and LMUexcellent, as well as the European Commission under the FET-Open project QNEMS (233992) is gratefully acknowledged.
We thank Darren R. Southworth for critically reading the manuscript.
\end{acknowledgments}

%

\end{document}